# The formation of labyrinths, spots and stripe patterns in a biochemical approach to cardiovascular calcification

**A Yochelis**[1,4], **Y Tintut**[1], **L L Demer**[1,2] and **A Garfinkel**[1,3]

[1]*Department of Medicine* (*Cardiology*), [2]*Department of Physiology*, [3]*Department of Physiological Science, University of California, Los Angeles, CA 90095*

**Abstract.** Calcification and mineralization are fundamental physiological processes, yet the mechanisms of calcification, in trabecular bone and in calcified lesions in atherosclerotic calcification, are unclear. Recently, it was shown in *in vitro* experiments that vascular-derived mesenchymal stem cells can display self-organized calcified patterns. These patterns were attributed to activator/inhibitor dynamics in the style of Turing, with bone morphogenetic protein 2 acting as an activator, and matrix GLA protein acting as an inhibitor. Motivated by this qualitative activator-inhibitor dynamics, we employ a prototype Gierer-Meinhardt model used in the context of activator-inhibitor based biological pattern formation. Through a detailed analysis in one and two spatial dimensions, we explore the pattern formation mechanisms of steady state patterns, including their dependence on initial conditions. These patterns range from localized holes to labyrinths and localized peaks, or in other words, from dense to sparse activator distributions (respectively). We believe that an understanding of the wide spectrum of activator-inhibitor patterns discussed here is prerequisite to their biochemical control. The mechanisms of pattern formation suggest therapeutic strategies applicable to bone formation in atherosclerotic lesions in arteries (where it is pathological) and to the regeneration of trabecular bone (recapitulating normal physiological development).

## Contents

---
[4] Author to whom any correspondence should be addressed. Email: yochelis@ucla.edu

## 1. Introduction

Cardiovascular calcification, in atherosclerosis or valvular stenosis, is considered one of the most notorious cardiac diseases [1]. The initial atherosclerotic lesion is formed as a soft cellular and fibrous mass (atheroma) growing within the artery wall. A myocardial infarction ("heart attack") occurs when the surface of the atheroma ruptures, exposing proteins that trigger clot formation in the blood. The clots then occlude the artery, preventing blood flow to the heart muscle. Little is known about the conditions under which rupture occurs. Indeed, there is even debate about whether calcium deposits mechanically stabilize or destabilize lesions [1]. Recently it was argued that "spotty" or "speckled" patterns of calcification carry the greatest risk for plaque rupture, as opposed to uniform deposits [1]. To advance the understanding of such pathology it is therefore essential to understand the mechanism determining the patterns formed by arterial calcification.

Arterial calcification is thought to be a recapitulation of embryonic bone formation by vascular mesenchymal stem cells (VMSCs) [2], under the control of bone morphogenetic protein 2 (BMP-2) [3-5]. In *in vitro* experiments, it was shown that cultures of vascular-derived mesenchymal stem cells differentiate and indeed spontaneously form calcified patterns out of a cellular monolayer [6]. Importantly, the morphology of the patterns was experimentally altered by applying external matrix GLA protein (MGP) (see figure 1); MGP molecules bind to active BMP-2, disabling their functionality. Thus they act as inhibitors. Assuming that the primary calcification mechanism is indeed preceded and governed by the interaction of two chemical morphogens, it is reasonable to test Turing's paradigm of morphogenesis to establish the chemical pre-patterning that shapes the pattern morphology.

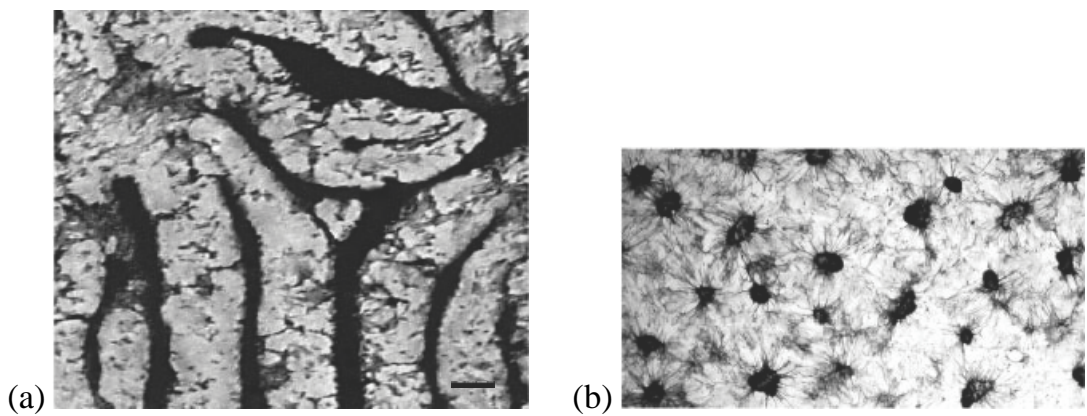

**Figure 1.** (a) A labyrinthine calcified (dark regions) pattern developed in vascular-derived mesenchymal stem cell culture; bar = 250 *μm*. (b) Spotted pattern after treatment with 40 *nM* exogenous MGP, which is an inhibitor of the activator BMP-2. (After [6])

Alan Turing, in his seminal work on morphogenesis [7], suggested that the formation of biological patterns can be understood by means of biochemistry, that is, in the reaction-diffusion framework. In this scenario, chemicals produced by cells interact as activators or inhibitors, and diffuse through the medium at distinct rates. This can create a symmetry breaking of the uniform concentrations, a mechanism that is often called a 'diffusion-driven instability' [8]. Since then, a number of morphogens have been identified [9] and linked to pattern development [6, 10-17]. These results suggest that an understanding of the dynamics of morphogenesis can give us a relatively simple way to understand and control biological development [16, 18-31].

The Turing approach assumes a decoupling between the biochemical processes and the biomass. However, the feasibility of the Turing paradigm in physiology faces the obstacle that the primary Turing instability is linear [32, 33]: the resulting pattern arises spontaneously and directly from a previously stable homogeneous condition. But the patterns that are typically observed are at *large* deviations from the critical conditions, and also on time scales *far* from the initial instability. Consequently, even if an initially homogenous biological system went through a Turing instability, it is impossible (using conventional methods) to experimentally demonstrate that fact.

In this article we discuss the role of reaction-diffusion mechanisms in promoting the mineralization patterns seen in VMSCs, as originally suggested in [6]. Using an activator-inhibitor model equation and analysis of pattern selection in the nonlinear regime, far from any initial bifurcation, Turing or otherwise, we explain previous *in vitro* VMSC culture experiments. Beyond that, we believe that a better description of the pattern formation phenomenon is prerequisite for biological experiments and future applications to stem cell calcification phenomena. The paper is organized as follows: in Section 2 we discuss a model equation. Next, in Section 3 we perform a one dimensional (1D) analysis to obtain the properties of periodic and localized steady states and discuss the pattern selection in the presence of multiple coexisting solutions. In Section 4, we use the latter results and secondary instabilities that operate in 2D to underline the mechanisms that lead to the formation of two dimensional patterns. We conclude and discuss the biomedical applications in Section 5.

## 2. Activator-inhibitor framework

The spatiotemporal biochemical activator-inhibitor dynamics in a VMSC culture were qualitatively modeled as follows: The monolayer of the VMSCs spontaneously expresses BMP-2 [34, 35] and its inhibitor matrix GLA protein (MGP) [36, 37], which diffuses more rapidly than the former [6]. With respect to local kinetics, it is assumed that BMP-2 obeys a saturated autocatalytic reaction [38] and directly promotes MGP production in a greater than linear fashion [39]. It is also assumed that both substances follow a first-order degradation. In addition, BMP-2 has a chemoattractant property [35] which is responsible for cell migration following the gradients of BMP-2. However, since: (*i*) the cell migration occurs at much slower time scales than the chemical diffusion, (*ii*) cell proliferation is relatively low [40], and (*iii*) neither BMP-2 nor MGP are consumed by the cells, we can neglect, to leading order, other contributions such as cell density, which indicate contributions of further, albeit minor, spatiotemporal changes of chemotactic distributions. In this approach the qualitative spatiotemporal distribution of the activator, BMP-2, emerges from interaction with the inhibitor, MGP, and thus the motion of the cells is driven by the active (MGP unbounded) BMP-2 gradients.

Following this description, we represent the active BMP-2 and MGP concentrations by $a(x,y)$ and $h(x,y)$, respectively. The calcified patterns in VMSCs cultures, developed from a monolayer in a dish size of the order of centimeters, i.e., implying a large aspect ratio system (large length vs. small height) that is quasi two dimensional. An activator-inhibitor model equation of such dynamics was proposed by Gierer and Meinhardt and has often been used to study biological pattern formation [29, 41]

$$\begin{aligned}
\frac{\partial a}{\partial t} &= D_a\left(\frac{\partial^2}{\partial x^2} + \frac{\partial^2}{\partial y^2}\right)a + \rho_a \frac{a^2 h^{-1}}{1+q^2 a^2} - \mu_a a + A, \\
\frac{\partial h}{\partial t} &= D_h\left(\frac{\partial^2}{\partial x^2} + \frac{\partial^2}{\partial y^2}\right)h + \rho_h a^2 - \mu_h h + H,
\end{aligned} \quad (1)$$

where $D_a, D_h$ are diffusion constants, $\rho_a, \rho_h$ are cross reaction coefficients, $\mu_a, \mu_h$ are degradation rates, $q$ is the saturation constant, and $A, H$ are source terms, respectively. We note that different variants and limits of this system have been discussed in [20, 26, 28, 30, 31, 42].

*2.1. Dimensionless forms, scaling, and parameters choice*

For our application, we set $A = 0$ due to the absence of activator source in experiments [6], and rewrite (1) in a dimensionless form, by introducing dimensionless variables $u = qa$, $v = \mu_a q h / \rho_a$, and scaling $t, \mathbf{x}$ by $\mu_a^{-1}, \sqrt{D_h / \mu_a}$, respectively:

$$\begin{aligned}\frac{\partial u}{\partial t} &= \frac{u^2 v^{-1}}{1+u^2} - u + D\left(\frac{\partial^2}{\partial x^2} + \frac{\partial^2}{\partial y^2}\right)u, \\ \frac{\partial v}{\partial t} &= Gu^2 - Ev + S + \left(\frac{\partial^2}{\partial x^2} + \frac{\partial^2}{\partial y^2}\right)v,\end{aligned} \quad (2)$$

where $D \equiv D_a / D_h, G \equiv \rho_h / \rho_a q, E \equiv \mu_h / \mu_a, S \equiv qH / \rho_a$. Biophysical parameter estimation from VMSC cultures [6] yields $D > 0 \sim O(10^{-2} - 10^{-3})$, $E > 1 \sim O(1)$, and $G > 0 \sim O(1)$. In what follows, we consider equation (2) for the parameter values

$$D = 0.005, \quad G = 1, \quad E = 2, \quad (3)$$

with $S$ as a *generalized* or *net* inhibition source, allowed to vary.

## 3. Steady-state solutions in 1D

*3.1. Uniform states and bistability*

Equations (2) and (3), admit three uniform solutions, one with trivial activator:

$$[u_0, v_0] = [0, S/2], \quad (4)$$

and two non-trivial

$$[u_\pm, v_\pm] = \left[\frac{\sqrt{\theta} \pm \sqrt{\eta}}{2}, \frac{\sqrt{\theta}\left(\sqrt{\theta} \pm \sqrt{\eta}\right)}{2 - (S-1)\sqrt{\theta} \pm \theta\sqrt{\eta}}\right], \quad (5)$$

where $\eta \equiv 4/\sqrt{\theta} + \phi$, $\theta \equiv \beta/3\rho + \rho/3 - 2(S+1)/3$, $\phi \equiv -\theta - 2(S+1)$, $\rho \equiv \sqrt[3]{\alpha/2 + \sqrt{\alpha^2/4 - \beta^3}}$ $\beta \equiv 1 + 14S + S^2$, and $\alpha \equiv 108 - 72S(S+1) + 2(S+1)^3$. These three uniform solutions, however, exist only for a certain range of values, $0 < S < S_{SN}$; in this range, both $(0, v_0)$ and $(u_+, v_+)$ are stable, defining a *bistability* region, as shown in figure 3. To leading order in $S$,

$$\begin{pmatrix}u_+ \\ v_+\end{pmatrix} - \begin{pmatrix}u_0 \\ v_0\end{pmatrix} \sim \begin{pmatrix}1 - \dfrac{S}{2} \\ 0\end{pmatrix} + O(S^2). \quad (6)$$

It is important to note that it is this bistability region, and in particular the large amplitude variations in the $u$ field, that makes possible the interspersed regions of high vs. vanishing cell densities observed in VMSC cultures [6].

*3.2. Periodic solutions and localized states*

In bistable systems, three primary types of stationary nonuniform solutions often arise [32, 33, 43]: (i) fronts connecting two uniform states (heteroclinic orbits in 1D physical space); (ii) periodic spatial patterns, also known as Turing patterns (that are limit cycles in 1D physical space); and (iii) localized states (homoclinic orbits in 1D physical space), *holes* or *peaks* superimposed on a background of a stable uniform state. The third type of state arises in systems with spatial reversibility $[u(-x), v(-x)] \to [u(x), v(x)]$ [43, 44]. In the following, we discuss the implications of periodic (ii) and localized (iii) solutions while a more general relation between solutions of type (i) and type (iii) is discussed in a companion paper [45].

First we address the temporal stability of the uniform states which is associated with non-uniform perturbations [32]

$$\begin{pmatrix} u \\ v \end{pmatrix} = \begin{pmatrix} u_* \\ v_* \end{pmatrix} + \varepsilon \begin{pmatrix} u_k \\ v_k \end{pmatrix} e^{\sigma t + ikx} + c.c. + O(\varepsilon^2) \qquad (7)$$

where $(u_*, v_*)$ is either $(0, v_0)$ or $(u_+, v_+)$, $\sigma$ is the perturbation growth rate, $k$ is the wavenumber, and *c.c.* stands for complex conjugate. The linear instability ($\varepsilon \ll 1$) of each one of the uniform states, is deduced from the dispersion relations: For $(u_*, v_*) = (0, v_0)$ the dispersion relations are given by

$$\sigma = -2 - k^2, \qquad (8)$$

and

$$\sigma = -1 - Dk^2, \qquad (9)$$

while for $(u_*, v_*) = (u_+, v_+)$, we obtain

$$\sigma_\pm = \frac{2u_+ - \gamma_+ \zeta^2 v_+}{2\zeta^2 v_+} \pm \frac{\sqrt{\gamma_-^2 \zeta^2 v_+^2 + 4\gamma_- \zeta^2 u_+ v_+ - 4u_+^2 v_+^2 (2u_+ \zeta^3 - 1)}}{2\zeta^2 v_+}, \qquad (10)$$

where $\gamma_\pm \equiv 2 \pm 1 + (1 \pm D)k^2, \zeta \equiv 1 + u_+^2$. Following (8), (9), and (3), the trivial state, $(0, v_0)$, is linearly stable to non-uniform perturbations ($\sigma < 0$). On the other hand, the nontrivial state $(u_+, v_+)$ exhibits a stationary finite wavenumber instability also known as the Turing instability [7, 32, 33], i.e., at $S = S_T \simeq 0.335$ there exists $\sigma(k = k_T) = 0$, $d\sigma/dk_T = 0$ and $d^2\sigma/dk_T^2 < 0$, where $k_T \simeq 4.55$ is the critical Turing wavenumber, as shown in figure 2. For $S > S_T$, the uniform state $(u_+, v_+)$ becomes unstable to a continuous band of wavenumbers.

Near the bifurcation onset, here Turing, wavenumber selection can be addressed using a weakly nonlinear analysis [32, 46]. However, numerical integration of (2) on a large domain ($L \gg 2\pi/k_T \equiv L_T \simeq 1.38$) with either periodic or reflecting boundary conditions, slightly above $S_T$ yields large amplitude periodic patterns, whose amplitudes approach $(0, v_0)$ and $(u_+, v_+)$. This behavior indicates the presence of a subcritical bifurcation, i.e., the branch of periodic states should bifurcate first, in direction $S < S_T$. Thus, in our case the weakly nonlinear analysis will not provide substantial insights into the nonlinear problem of pattern selection.

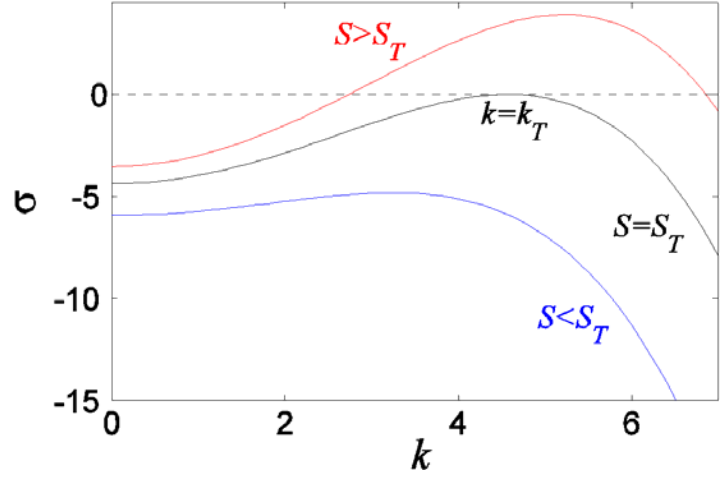

**Figure 2.** The growth rates, $\sigma = \sigma_+(k)$, of periodic perturbations, according to Eq. (10). The critical Turing mode $k = k_T \simeq 4.55$, is obtained at $S = S_T \simeq 0.335$ (middle), while other curves correspond to (bottom) $S = 0.2$ and (top) $S = 0.4$.

To understand the pattern selection mechanism, we exploit first the method of *spatial dynamics* [44], which was found to be an effective method exploring the multiplicity of steady state solutions [43, 47-52]. To do so, we set $\partial_t u = \partial_t v = 0$ and rewrite (2) as a set of four first order ordinary equations

$$\begin{aligned}
\partial_x u &= -h, \\
\partial_x h &= \frac{1}{D}\left(\frac{u^2 v^{-1}}{1+u^2} - u\right), \\
\partial_x v &= -w, \\
\partial_x w &= Gu^2 - Ev + S,
\end{aligned} \qquad (11)$$

where space is now treated as a time-like variable. Next, we compute the branches of steady states that bifurcate from the Turing onset which corresponds, in this specialized case (eq. (11)), to a reversible Hopf bifurcation [44], via a numerical continuation method [53]. Indeed, the instability of the $(u,h,v,w) = (u_+, 0, v_+, 0)$ state gives rise to a subcritical bifurcation of periodic states, accompanied by a subcritical bifurcation of spatial groups consisting of a finite number of localized hole solutions [45]. Since the bifurcation is from a nontrivial state, terms 'beyond all order' in the asymptotic analysis select only two families of even parity groups, distinguished by a phase shift of $\pi$ and corresponding respectively, to odd and even number of hole groups [48, 54]. Notably, the simultaneous bifurcations of both periodic Turing, $L = L_T$, and groups of localized solutions, $L \to \infty$, are known to arise in variational systems [47-51]. The bifurcation diagram of these solutions is represented in figure 3, where we use the norm

$$N = \sqrt{L^{-1}\int_0^L dx\left\{u^2 + v^2 + \left(\partial_x u\right)^2 + \left(\partial_x v\right)^2\right\}}, \tag{12}$$

to distinguish among solutions, where $L$ is the spatial period.

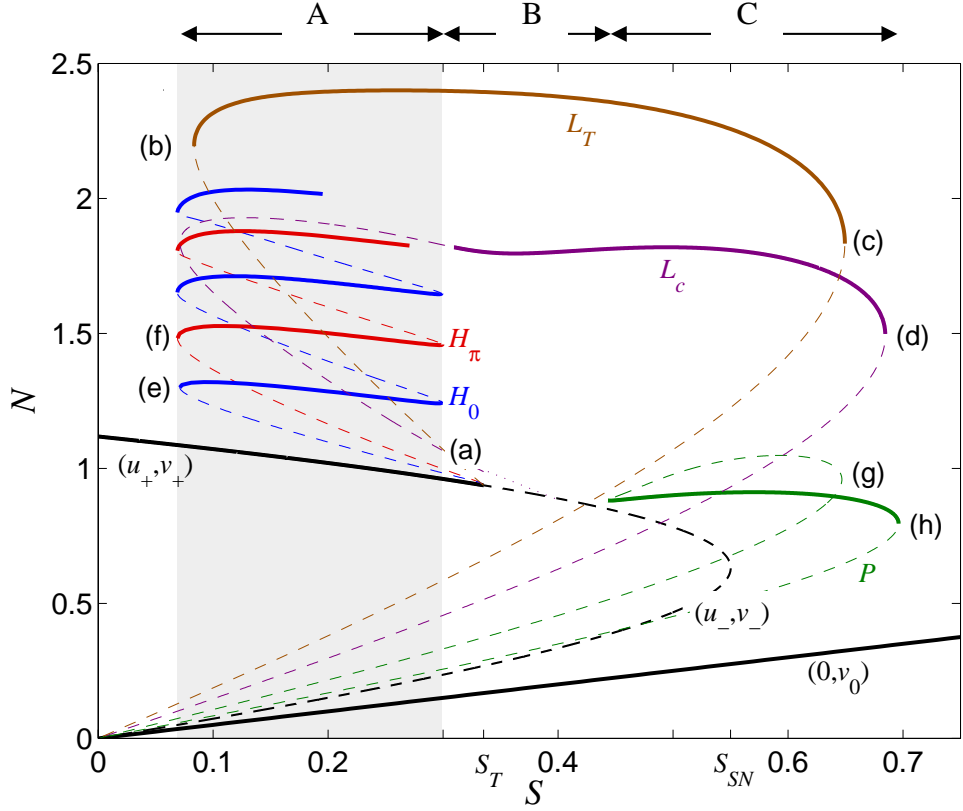

**Figure 3.** Bifurcation diagram showing coexisting branches of uniform solutions, $(0,v_0),(u_\pm,v_\pm)$, periodic solutions with distinct periods, $L=L_T$, $L=L_c$, and localized solutions in the form of holes $H_{0,\pi}$ and peaks $P$. The solutions are plotted in terms of $N$, see equation (12), as a function of $S$. All the solutions were obtained by integration of (11) via AUTO [53] on periodic domains, and their stability was determined by a numerical eigenvalue method using (2). Solid lines mark stable portions of the branch; the shaded region represents the pinning regime of the spatially bounded hole states. The $H_0$ and $H_\pi$ are branches of bounded states with a respective odd and even number of holes, each stable branch associated with a distinct number of holes which increases with $N$. The location of the saddle node of uniform states $(u_\pm,v_\pm)$ is indicated by $S=S_{SN}$. The profiles along the stable portions of each branch agree with the results of direct numerical integration of (2) with periodic boundary conditions. (a-h) Mark the distinct solutions along the stable and unstable branches and are presented in figure 4.

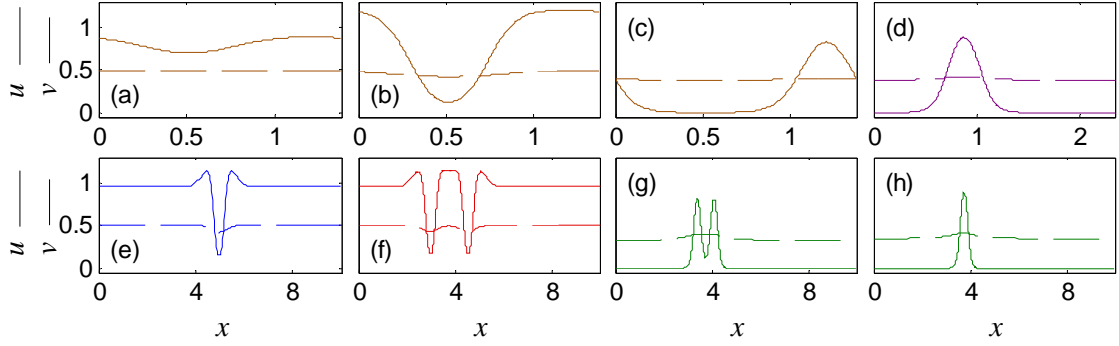

**Figure 4.** (a-h) Steady state profiles of $u$ (solid lines) and $v$ (dashed lines) computed in periodic domains at locations indicated in figure 3. Period length: $L_T \equiv 2\pi/k_T \simeq 1.38, L_c \simeq 2.34, H_{0,\pi} = P = 10$, where $k_T$ is the Turing mode obtained at $S = S_T$. We verified that both holes, $H_{0,\pi}$, and peaks, $P$, are indeed localized solutions by increasing the period to $L = 100$. Note that (a) marks the profile on the unstable branch of $L_T$ in vicinity of the Turing bifurcation point.

Consequently, the periodic Turing state $L = L_T$ (see profile (a) in figure 4), and the hole states $H_{0,\pi}$, appear as unstable small amplitude solutions around the uniform $(u_+, 0, v_+, 0)$ state and become large amplitude with respect to the activator (profile (b), (e), (f) in figure 4), at the first saddle nodes (see figure 3). The odd and even hole branches form a pinning region (see the shaded region in figure 3), where higher branches indicate increasing number of holes [45]. In variational systems, these oscillations cause an effective broadening of the Maxwell point [47-51] at which a heteroclinic orbit in space (Pomeau front) between the uniform and periodic solutions is time independent [55] so that the finite group of bounded holes is a connection between two heteroclinic orbits in space, i.e., a *heteroclinic cycle*.

The primary periodic Turing solutions $L = L_T$, gain stability (also in large periodic domains $L = nL_T$, where $n > 1$) at the left saddle node, remain stable till the right saddle node and reconnect to the $u = v = 0$ state at $S = 0$, as shown in figure 3. However, numerical integration above the rightmost saddle node, surprisingly show that periodic patterns may still persist. The origin of this behavior can be understood by examination of the secondary bifurcating wavenumbers. As $S$ increases beyond $S_T$, the $(u_+, 0, v_+, 0)$ state loses stability to additional wavenumbers (see figure 2). To demonstrate the impact of small wavenumbers (due to our interest in large domains $L > L_T$), we have followed only the solution with the period, $L = L_c \simeq 2.34 < 2L_T$, that bifurcates around $S = 0.4$. We note that there exist an infinite number of unstable wavenumbers, which for simplicity are not shown in figure 3. These solutions also bifurcate subcritically and reconnect also to the origin, but unlike the $L = L_T$ solutions, these solutions gain temporal stability after the left saddle node (see figure 3). Notably, all the periodic solutions near the left saddle nodes are an inverse image (with respect to the activator field) of the solutions (with same period) near the right saddle nodes, that is, hole-like vs. peak-like. The solutions arise first as hole-like states due to the Turing instability of the uniform nontrivial state $(u_+, 0, v_+, 0)$. This is demonstrated for the primary $L = L_T$ state, see profiles (b) and (c) in figure 4, respectively. The stable branches form a finite region at which periodic solutions can form either spontaneously, around $(u_+, 0, v_+, 0)$ for $S_T < S < S_{SN}$, or by large amplitude perturbations above $(0, v_0)$ for $S > S_{SN}$ (see figure 3 for details). We propose that it is this property that makes possible the chemotactic gradients that control cell migrations observed in VMSC cultures [6].

As already noted, all the periodic solution branches terminate at the onset of the transcritical bifurcation, $S = 0$. The successive structure of the saddle nodes of periodic solutions above $S = S_{SN}$, suggest that localized peak states, with period $L \to \infty$, can also be present. Indeed such homoclinic orbits in space, $P$, do appear but they do not bifurcate from the nontrivial branch like the periodic states. Nevertheless, by prolongation of the periodic solutions' period in the vicinity of $S = 0$, we computed two distinct states, unstable (figure 4g) and stable (figure 4h), which biasymptote to $(0,0,v_0,0)$. However, unlike the possible multiplicity of the stable groups of hole solutions, only a single peak can stabilize (profile (h) in figure 4) since the double peak solution is unstable (profile (g) in figure 4), implying a *repulsive* interaction between two neighboring peaks. While the single type of peak has been observed in numerical integrations of (2) before [41], their origin is unclear due to the linear temporal stability of the $(u,v) = (0, v_0)$ state (see equations (8) and (9)). To address this issue we explore the asymptotic approach of peak states to $(0,0,v_0,0)$, as $x \to \pm\infty$, which is given by

$$\begin{pmatrix} u \\ h \\ v \\ w \end{pmatrix} - \begin{pmatrix} 0 \\ 0 \\ v_0 \\ 0 \end{pmatrix} \propto e^{\lambda x}, \tag{13}$$

where the four spatial eigenvalues are real, $\lambda = \pm\sqrt{E} = \pm\sqrt{2}$ and $\lambda = \pm\sqrt{1/D} = \pm\sqrt{200}$. In spatially reversible systems, the formal bifurcation of localized states with monotonic tails can be either as a small amplitude (spatial) bifurcation from a uniform state [52] or by a nucleation from a global heteroclinic bifurcation, connecting in space two uniform states [45]; the latter is absent in this parameter set. In the small amplitude bifurcation case, the four real spatial eigenvalues at the bifurcation point should correspond to $(\lambda_1, \lambda_2, \lambda_3, \lambda_4) = (0, 0, +\mu, -\mu)$ and for $S > 0$ the situation should read $(\lambda_1, \lambda_2, \lambda_3, \lambda_4) = (+\nu, -\nu, +\mu, -\mu)$ [44, 52], where $\mu, \nu > 0$ and real. This scenario allows a transverse intersection of two manifolds: a two-dimensional stable manifold and two-dimensional unstable manifold, and thus the possible formation of homoclinic orbits (in space) with monotonic tails. Since the spatial eigenvalues are independent of $S$, the localized $P$ states whose eigenvalues are always of the form $(-\sqrt{E}, +\sqrt{E}, -\sqrt{1/D}, +\sqrt{1/D})$, can not formally bifurcate from the trivial $(0,0,v_0,0)$ state as small amplitude solutions, although their amplitude at $S = 0$ is zero, as shown in figure 3. This should not come as a surprise since the bifurcation condition $E \to 0$, simultaneously requires $S \to 0$, due to the blow up of the trivial state $(0,0,v_0,0)$, where $v_0 = S/E$.

To qualitatively understand the peculiar connection between the periodic states and the origin of localized peaks at $S = 0$, we use a projection of the four dimensional phase space onto the 2D subspace $u$ vs. $\partial_x v$ [44, 50]. As already mentioned, in this phase space the periodic states are limit cycles; in the following we discuss only $L = L_T$. This limit cycle persists above $S = S_{SN}$, and in the absence of uniform states bistability approaches the $(0,0,v_0,0)$ state as $S$ is increased (see figure 5a). At the rightmost saddle node, the limit cycle within numerical precision crosses the trivial state $(0,0,v_0,0)$, however, the formation of a heteroclinic bifurcation between the periodic and the trivial state is impossible due to the real forms of the spatial eigenvalues [44, 45, 48]. This behavior, on the other hand, implies a transition to a homoclinic orbit $P$ to the fixed point $(0,0,v_0,0)$. However, without the existence of a heteroclinic bifurcation between the two uniform states [45] a different mechanism should be considered for this special situation in which no distinction between two orbits can be made. This happens exactly at $S = 0$, where uniform, periodic and peak states are simply zero; figure 5b shows the vanishing amplitude of the periodic state. Therefore the transcritical bifurcation of uniform states also serves as an effective bifurcation point for the localized peak solutions and a reason for termination of the periodic states, as demonstrated in figure 3.

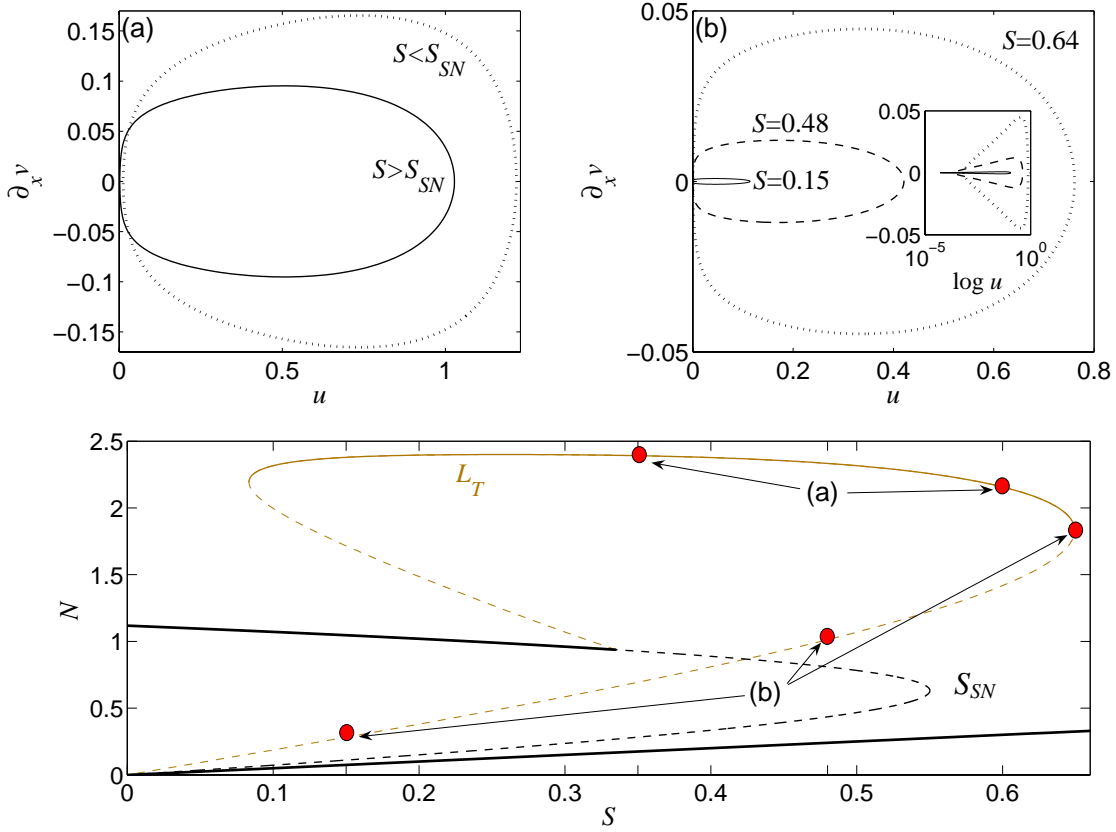

**Figure 5.** Upper Panel: Projections of the periodic Turing solution in a phase portrait $u$ vs. $\partial_x v$, along (a) stable and (b) unstable branches as depicted in the bottom panel. The inset in (b) shows the plane as a function of $\log u$, to emphasize the limit cycle approach towards the fixed point $(u,h,v,w) = (0,0,v_0,0)$ as $S \to 0$.

*3.3. Wavenumber selection in presence of multiple solutions*

As has been described, the nonlinear regime exhibits the coexistence of multiple stable solutions, making it hard to foresee the wavenumber selection and the sensitivity to initial conditions; both determine the basin of attraction of the final states. To understand the mechanism, we focus here on three regions distinguished by coexisting solutions (see figure 3): (A) bounded holes and periodic states; (B) periodic states; and (C) periodic and isolated peak states.

Since in the pinning region (A), fronts between localized and periodic solutions are stationary, we refer the reader to [48], and discuss in the following only regions (B) and (C). In the linearly unstable regime $S_T < S < S_{SN}$, infinitesimal random perturbations around the $(u_+, v_+)$ state grow and form periodic patterns with periodicity $L = L_T$, due to the fastest growth of the critical Turing mode. However, other initial conditions give different results. In both regions (B) and (C), a finite amplitude spatially nonuniform initial condition that, for example, features two different length scales in two parts of the domain, results in relaxation by phase diffusion [56, 57]: the asymptotic periodicity is a *compromise* value between two initial periodic states, implying dispersion of the front separating two domains with different periodicities rather than invasion or phase slips, see figure 6. When we initiated a domain with two different stable solutions coexisting in regions (B) and (C), it then evolved into a single final pattern, whose length scales were intermediate, which are also stable solutions of (2) (figures 6a and 6b, respectively).

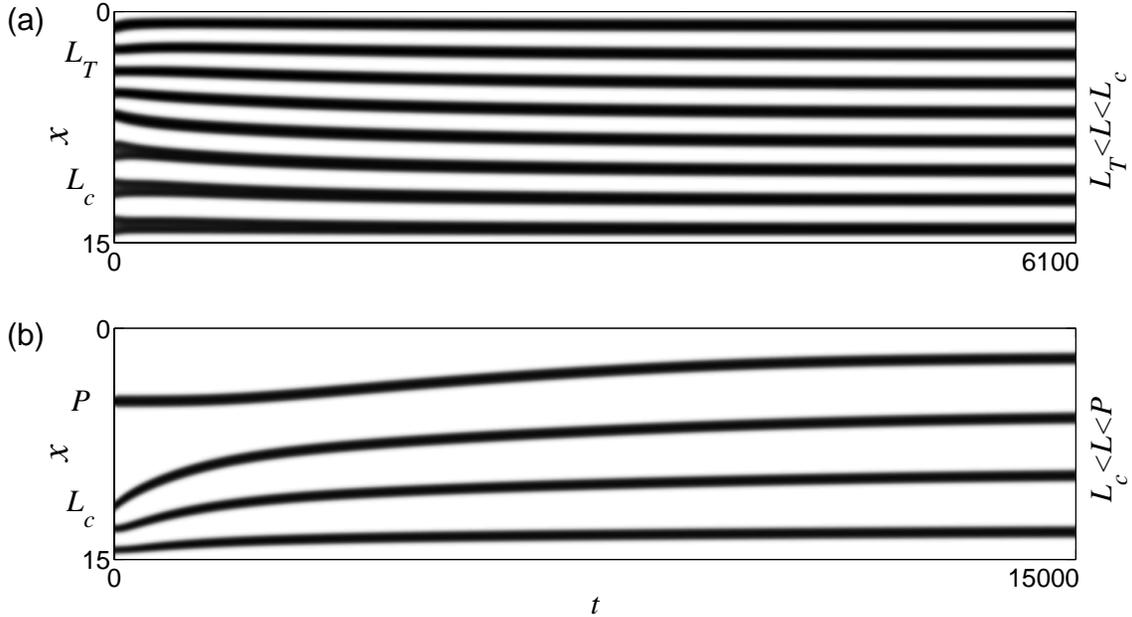

**Figure 6.** Space-time plots showing the temporal evolution of an initial condition that is a combination of (a) $L_T \simeq 1.38$ and $L_c \simeq 2.34$, and (b) $L_c$ and $P$ (as indicated at $t = 0$). Equation (2) was integrated in 1D with Neumann boundary conditions; a grey-scale map shows the $u$ field. Parameters: (a) $S = 0.35$, (b) $S = 0.5$.

## 4. Mechanisms of pattern formation in 2D

It is well-known that reaction diffusion systems can exhibit transverse and curvature-induced instabilities in 2D [32, 33, 58], so that labyrinthine or spotted patterns can form respectively, through secondary zigzag or varicose instabilities of stripes [59] or through a transverse instability of axisymmetric spots [60]. Thus, we have performed an extended numerical investigation to obtain the secondary pattern formation mechanisms of the diverse patterns.

First, we have numerically calculated the regions of zigzag, varicose and curvature instabilities of periodic stripes, isolated stripes, and isolated spots, respectively. The distinct instability types and their ranges indicated in figure 7 and respectively shown in figure 8. We have exploited the 1D periodic and localized solutions obtained above to construct the respective periodic (figure 8a) and isolated stripe (figure 8a) solutions in 2D while the spot (figure 8c) solutions were obtained by a relaxation from an axisymmetric construction based on the initial localized peak or hole state size. We do not show the transient evolutions from spots toward labyrinthine patterns in the regime $S_T < S < S_{SN}$, and refer the reader to discussions of similar phenomena in the context of isolated stripes [28] and isolated spots [60].

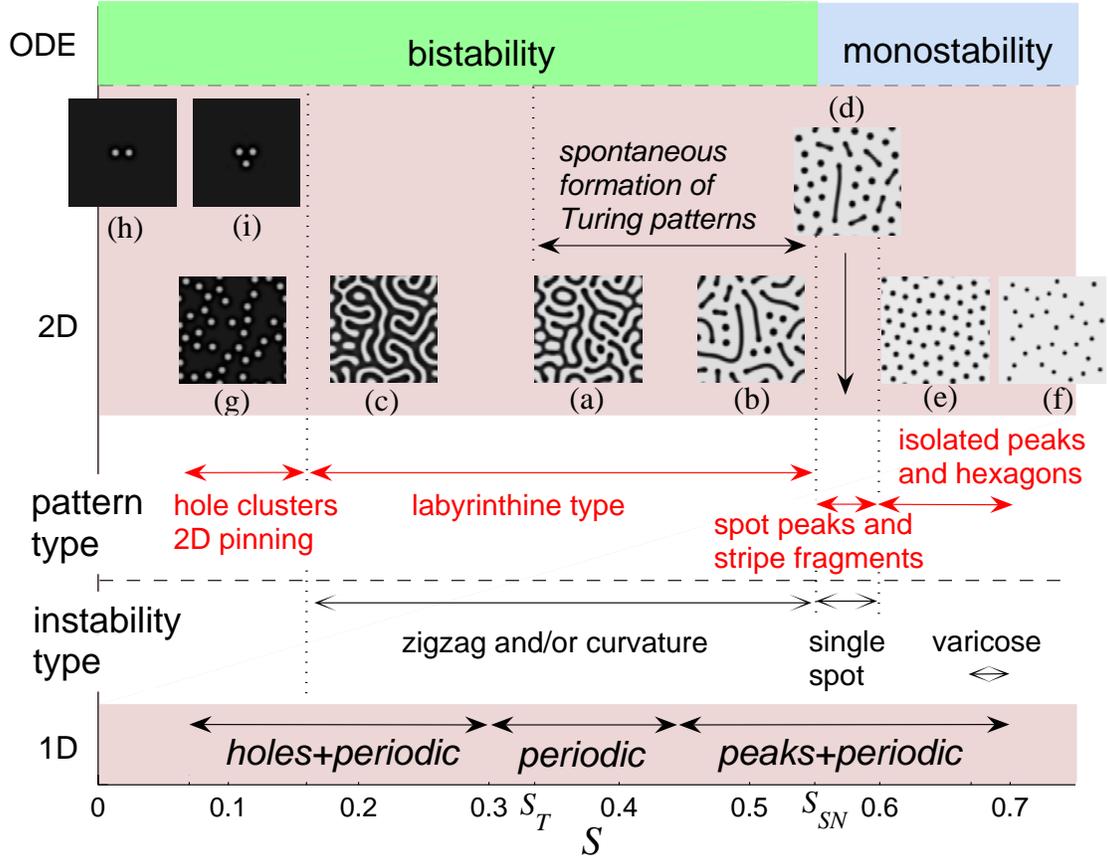

**Figure 7.** Diagram showing the instabilities and the corresponding patterns forms as a function of $S$, where 'zigzag' corresponds to instability of stripes while 'curvature' is used in the context of spots; both instabilities lead to formation of labyrinthine-type patterns. The distinct instabilities are respectively demonstrated in figure 8. The two dimensional patterns were obtained via numerical integration of (2) with periodic boundary conditions on a $x = y = [0,15]$ physical domain; dark regions correspond to higher values of the $u$ field. Labyrinthine patterns (a) were generated from infinitesimal random perturbations around $(u_+, v_+)$, and used as initial conditions for (b-g). The spotted patterns (e-g) are not asymptotic; they are shown for $t = 3000$. The repulsive interactions between the peak spots in (e-f) are very weak, however the final state eventually approaches hexagonal symmetry, while in (g) there is deformation towards hole clusters rather than to hexagonal symmetry. Localized patterns (h-i) are examples of the pinning phenomenon that persist in two dimensions. Parameters: (a) $S = 0.35$, (b) $S = 0.53$, (c) $S = 0.2$, (d) $S = 0.58$, (e) $S = 0.62$, (f) $S = 0.69$, (g-i) $S = 0.1$.

One of the primary interests in the experimental context is the control of the chemical pre-pattern under various concentrations of the inhibitor source. Thus, to demonstrate the effect of these instabilities on the asymptotic pattern formation, we have at first integrated equation (2) in 2D above the Turing onset, starting from infinitesimal random perturbations around the uniform $(u_+, v_+)$ state; as expected due to the zigzag instability (see figure 8a), a labyrinthine pattern was formed (see inset (a) in figure 7). Next, we used this pattern as an initial state for other values of $S$, the results are shown in insets (b-g) in figure 7.

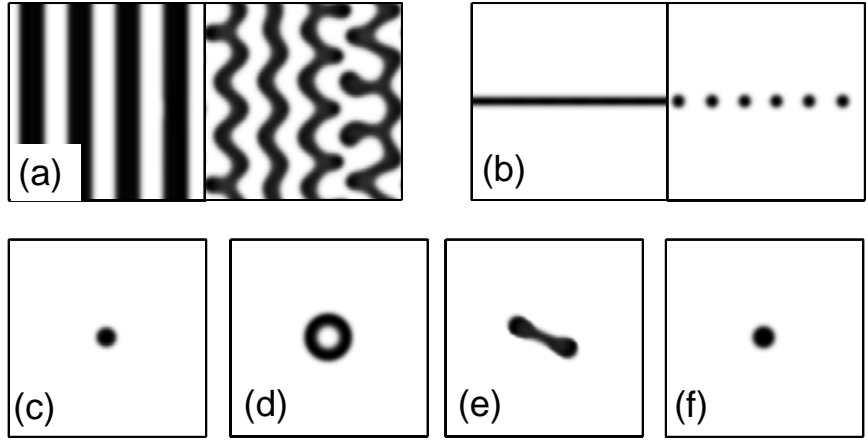

**Figure 8.** Numerical integration of (2) on a 2D domain, $x = y = [0,10]$, with periodic boundaries; dark regions correspond to higher values of the $u$ field. (a) Zigzag instabilities of an initial stripe state with $L = L_c \simeq 2.34$, times from left to right: $t = 0, 828$. (b) Varicose (breakup) instability of a localized stripe, times from left to right: $t = 0, 400$. (d,e) Instability and (f) stability of an initial (slightly perturbed) axisymmetric localized spot (c); frames (d) and (e) are transients. Parameters: (a) $S = 0.35$, (b) $S = 0.67$, (c) initial condition of isolated spot, (d) $S = 0.5$ at $t = 120$, (e) $S = 0.55$ at $t = 2000$, (f) $S = 0.6$ at $t = 3000$.

The domain of labyrinthine-type patterns ranges from the bistability onset at $S = S_{SN}$ (inset (b) in figure 7) to the 2D pinning region (inset (c) in figure 7), however only for $S_T < S < S_{SN}$ do they form spontaneously, due to the Turing instability of the uniform $(u_+, v_+)$ state. In a small region above $S_{SN}$, single spots of peak type are still unstable (see figure 8e), so that straight stripe fragments may form. In case of initial labyrinthine pattern (inset (a) in figure 7) this leads to the formation of periodic states consisting of both spots and stripe fragments (inset (d) in figure 7).

As a single spot stabilizes (see figure 8f), the pattern becomes ultimately spotted. However, according to figure 3, the pattern may admit distinct periodicities under other initial conditions; direct numerical 2D computations support this prediction, as insets (e-f) in figure 7 show. We note that both states (e-f) are not asymptotic and approach asymptotically to a hexagonal symmetry, however, since the interaction between localized states is weakly repulsive, this process is very slow. For small $S$ values (that is, in the 2D pinning region [61, 62]), we obtain an image inverse to peak spots, that is, distributed *hole spots* (inset (g) in figure 7). In the absence of repulsive interactions between neighboring holes, see insets (h-i) in figure 7, the hole spots support coexisting *clusters*, as shown in inset (g) in figure 7.

## 5. Conclusions

In this paper, we theoretically studied the mechanisms of pattern formation that give rise to many steady state patterns in a biologically oriented activator-inhibitor model [29, 41]. We chose a version with an inhibition source term, and a saturated autocatalytic local kinetics. These resulted in a bistability regime of uniform solutions, distinguished by finite and zero activator concentrations (see equation (6)). While different forms of the Gierer-Meinhardt model have been studied, and a number of interesting qualitative phenomena have been found [20, 26, 28, 30, 42], we present here a generalized and detailed view of the multiple periodic and localized states that occur in this model, as well as their basins of attraction. In particular, we show that the majority of patterns for high inhibitor source values form subcritically, that is, by finite amplitude perturbations above the trivial state $(0, v_0)$, as shown in figures 3 and 7. First, we exploited the powerful tool of spatial dynamics coupled with numerical continuation, and incorporated with temporal stability, to identify the primary instabilities and the possible stable and unstable 1D solution branches as a function of the inhibition source, $S$ (see figure 3). We used this analysis to calculate the secondary instabilities and the parameter regions in which 2D patterns, such as labyrinths, mixtures of spots and stripe fragments, and bounded vs. isolated spots, can form (see figure 7). The knowledge of the nonlinear pattern selection mechanisms *far* from the Turing onset is therefore an important missing link in the context of robustness and pattern control, which up to now was considered as one of the weak points of the activator-inhibitor approach in biological pattern formation [63].

The qualitative form of the model, and in particular the parameter values, were made to keep fidelity with our main application, which is the formation of calcified patterns by vascular mesenchymal stem cells (VMSCs) [6]. These are the cells that are thought to form bone in atherosclerotic tissue, the primary process that has been linked to atherosclerotic calcification [2]. The approach we employ is associated with the primary symmetry breaking mechanism of a cellular monolayer. This assumption is justified by the fact that cells do not consume the morphogens, allowing a qualitative decoupling between the biochemistry and the biomass. Thus, we have excluded from this model any higher order contributions, such as the contributions of changing cell densities, which may support chemotaxis-based mechanisms [30]. Within these lines, cell density is expected to have a minor role in the initial qualitative pattern selection, and only quantitatively modify the pattern architecture at later time stages. The extension of the Gierer-Meinhardt model to include the contributions of cell density is beyond the scope of this paper, and will be discussed elsewhere.

Importantly, a spontaneous formation of labyrinthine and spotted patterns (with roughly hexagonal symmetry) has indeed been observed in *in vitro* experiments in VMSC preparations, as shown in figure 1, and discussed in more detail in [6]. In these experiments the inhibition source was altered by external addition of MGP, the inhibitor of the activator BMP-2. As in our analysis, labyrinthine patterns have been found at lower concentrations of added MGP (corresponding to lower $S$ values) while at higher MGP concentrations (corresponding to higher $S$ values) periodic spotted patterns were developed. The good qualitative agreement between the theoretical analysis and the experimental observations suggests a novel strategy for the biochemical control of calcified patterns, via the framework of activator-inhibitor dynamics.

There are several potential applications of this strategy. First, as noted above, the spotted patterns of calcification seen in culture dishes are considered a good model for the formation of spotty atherosclerotic calcification in humans [1]. If these calcified deposits become confluent, it has been conjectured that the resulting solid mass is expected to mechanically stabilize the adjacent lesion, while if the calcification remains 'spotty', there is an increased risk of rupture and myocardial infarction [1]. We believe that the theoretical prediction of patterns will shed light on design of future experiments and allow also identification of the proper biochemical conditions for spot nucleation phenomenon (peak states), which according to the theory can not be observed spontaneously and require specific initial conditions.

A second application is to the formation of bone tissue (by the same types of cells). Trabecular bone has a dense labyrinthine and hole architecture, presumably formed by the same morphogenetic processes, so our analysis may also be relevant to biochemical based methods of bone regeneration.

**Acknowledgments**

We thank Edgar Knobloch, John Burke, and Kristina Boström for helpful discussions and the anonymous referees for their valuable comments. This work was supported by NIH Grant HL0881202 and by NIH/NHLBI Grant P01 HL078931.

**References**

[1] Abedin M, Tintut Y and Demer L L 2004 *Arterioscler Thromb Vasc Biol* **24** 1161-70
[2] Abedin M, Tintut Y and Demer L L 2004 *Circ Res* **95** 671-6
[3] Tintut Y, *et al.* 2003 *Circulation* **108** 2505-10
[4] Hruska K A, Mathew S and Saab G 2005 *Circ Res* **97** 105-14
[5] Shao J S, Cai J and Towler D A 2006 *Arterioscler Thromb Vasc Biol* **26** 1423-30
[6] Garfinkel A, *et al.* 2004 *Proc Natl Acad Sci USA* **101** 9247-50
[7] Turing A M 1952 *Philos Trans R Soc London, Ser B* **237** 37-72
[8] Maini P K, Painter K J and Chau H N P 1997 *J Chem Soc-Faraday Trans* **93** 3601-10
[9] Lander A D 2007 *Cell* **128** 245-56
[10] Sick S, Reinker S, Timmer J and Schlake T 2006 *Science* **314** 1447-50
[11] Morris H R, *et al.* 1987 *Nature* **328** 811-4
[12] Chen Y and Schier A F 2001 *Nature* **411** 607-10
[13] Harris M P, *et al.* 2005 *Proc Natl Acad Sci USA* **102** 11734-9
[14] Painter K J, Maini P K and Othmer H G 1999 *Proc Natl Acad Sci USA* **96** 5549-54
[15] Yamaguchi M, Yoshimoto E and Kondo S 2007 *Proc Natl Acad Sci USA* **104** 4790-3
[16] Liu R T, Liaw S S and Maini P K 2006 *Phys Rev E* **74** 011914
[17] Yu M, Wu P, Widelitz R B and Chuong C M 2002 *Nature* **420** 308-12
[18] Hunding A and Sorensen P G 1988 *J Math Biol* **26** 27-39
[19] Murray J D and Oster G F 1984 *IMA J Math Appl Med Biol* **1** 51-75
[20] Arcuri P and Murray J D 1986 *J Math Biol* **24** 141-65
[21] Dillon R, Maini P K and Othmer H G 1994 *J Math Biol* **32** 345-93
[22] Klein C T 1998 *J Theor Biol* **194** 263-74
[23] Crampin E J, Gaffney E A and Maini P K 1999 *Bull Math Biol* **61** 1092-120
[24] Barrio R A, Varea C, Aragon J L and Maini P K 1999 *Bull Math Biol* **61** 483-505
[25] Hunding A 1987 *J Math Biol* **25** 109-21
[26] Ward M J, *et al.* 2002 *SIAM J Appl Math* **62** 1297-328
[27] Mimura M A and Nishiura Y 1979 *J Math Biol* **7** 243-63
[28] Kolokolnikov T, Sun W, Ward M and Wei J 2006 *SIAM J Appl Dyn Syst* **5** 313-63
[29] Gierer A and Meinhardt H 1972 *Kybernetik* **12** 30-9
[30] Murray J D 2002 *Mathematical Biology* (New York: Springer)
[31] Hillen T 2007 *SIAM Rev* **49** 35-51
[32] Cross M C and Hohenberg P C 1993 *Rev Mod Phys* **65** 851-1112
[33] Pismen L M 2006 *Patterns and Interfaces in Dissipative Dynamics* (Berlin: Springer-Verlag)
[34] Boström K, *et al.* 1993 *J Clin Invest* **91** 1800-9
[35] Willette R N, *et al.* 1999 *J Vas Res* **36** 120-5
[36] Boström K, *et al.* 2001 *J Biol Chem* **276** 14044-52
[37] Zebboudj A F, Imura M and Boström K 2002 *J Biol Chem* **277** 4388-94
[38] Ghosh-Choudhury N, *et al.* 2001 *Biochem Biophys Res Commun* **286** 101-8
[39] Zebboudj A F, Shin V and Boström K 2003 *J Cell Biochem* **90** 756-65
[40] Bennett R L, Navab M, Demer L L and Fogelman A M 1993 *Arterioscler Thromb* **13** 360-6
[41] Koch A J and Meinhardt H 1994 *Rev Mod Phys* **66** 1481-507
[42] Doelman A and van der Ploeg H 2002 *SIAM J Appl Dyn Syst* **1** 65-104
[43] Knobloch E 2008 *Nonlinearity* **21** T45-T60
[44] Champneys A R 1998 *Physica D* **112** 158-86
[45] Yochelis A and Garfinkel A 2008 *Phys Rev E* **77** R035204
[46] Hoyle R B 2006 *Pattern Formation: An Introduction to Methods* (Cambridge: University Press)
[47] Woods P D and Champneys A R 1999 *Physica D* **129** 147-70
[48] Burke J and Knobloch E 2006 *Phys Rev E* **73** 056211
[49] Burke J and Knobloch E 2007 *Phys Lett A* **360** 681-8

[50] Gomila D, Scroggie A J and Firth W J 2007 *Physica D* **227** 70-7
[51] Coullet P 2002 *Int J Bifurcation Chaos* **12** 2445-57
[52] Yochelis A, Burke J and Knobloch E 2006 *Phys Rev Lett* **97** 254510
[53] E. Doedel *et al.* AUTO2000: Continuation and bifurcation software for ordinary differential equations (with HomCont). Available from: http://indy.cs.concordia.ca/auto/
[54] Kozyreff G and Chapman S J 2006 *Phys Rev Lett* **97** 44502
[55] Pomeau Y 1986 *Physica D* **23** 3-11
[56] Paap H G and Riecke H 1991 *Phys Fluids A* **3** 1519-32
[57] Pomeau Y and Manneville P 1979 *J Phys Lett* **40** L609-L12
[58] Burke J and Knobloch E 2007 *Chaos* **17** 37102
[59] Hirschberg P and Knobloch E 1993 *Chaos* **3** 713-21
[60] Kolokolnikov T and Tlidi M 2007 *Phys Rev Lett* **98** 188303
[61] Lloyd D J B. Localised Solutions of Partial Differential Equations [Ph.D. thesis]: University of Bristol; 2005.
[62] Lloyd D J B, Sandstede B, Avitabile D and Champneys A R unpublished
[63] Brenner M P, Levitov L S and Budrene E O 1998 *Biophys J* **74** 1677-93